%% file: main.tex
\documentclass[preprint,12pt, a4paper]{elsarticle}

\usepackage{amssymb}
\newcommand{\code}[1]{\texttt{#1}}

\usepackage[htt]{hyphenat}
\usepackage{adjustbox}
\usepackage{courier}
\usepackage{listings}
\usepackage{caption}
\usepackage{color}
\usepackage{xcolor}
\definecolor{light-gray}{gray}{0.975}
\DeclareCaptionFormat{listing}{\parbox{\textwidth}{#1#2#3}}
\usepackage{hyperref}
\usepackage[nameinlink]{cleveref}

\usepackage{float}
\restylefloat{table}

\journal{SoftwareX}

\begin{document}
\input{frontmatter}
\newpage
\section*{Required Metadata}
\label{section:metadata}

\begin{table}[H]
\begin{tabular}{|l|p{6.5cm}|p{6.5cm}|}
\hline
C1 & Current code version & 0.6.0\\
\hline
C2 & Permanent link to code/repository used for this code version & \href{https://github.com/pesummary/pesummary/tree/v0.6.0}{https://github.com/pesummary/\linebreak pesummary/tree/v0.6.0} \\
\hline
C3 & Legal Code License   & MIT \\
\hline
C4 & Code versioning system used & git \\
\hline
C5 & Software code languages, tools, and services used & Python, Javascript, HTML, CSS, Bootstrap, Unix/MacOS \\
\hline
C6 & Compilation requirements, operating environments \& dependencies & numpy$\geq 1.15.4$, h5py, matplotlib, seaborn, statsmodels, corner, tables, deepdish, pandas, pygments, astropy$\geq 3.2.3$, lalsuite$\geq 6.70.0$, ligo-gracedb, configparser, gwpy, plotly, tqdm$\geq 4.44.0$
\\
\hline
C7 & If available Link to developer documentation/manual & See \href{https://lscsoft.docs.ligo.org/pesummary}{https://lscsoft.docs.ligo.org/\linebreak pesummary} \\
\hline
C8 & Support email for questions & \href{mailto:charlie.hoy@ligo.org}{charlie.hoy@ligo.org} \\
\hline
\end{tabular}
\caption{Code metadata (mandatory)}
\label{} 
\end{table}

\input{introduction}
\input{software}
\input{examples}
\input{impact}
\input{conclusion}

\input{conflict}

\input{acknowledgements}




\bibliographystyle{elsarticle-num} 
\bibliography{main}

\end{document}

%% file: frontmatter.tex
\begin{frontmatter}



\title{{\sc{PESummary}}: the code agnostic Parameter Estimation Summary page builder}

\author[label2]{Charlie Hoy\corref{cor1}}
\cortext[cor1]{Corresponding Author}
\ead{charlie.hoy@ligo.org}
\author[label2]{Vivien Raymond}
\address[label2]{Cardiff University, Cardiff CF24 3AA, UK}

\begin{abstract}
PESummary is a Python software package for processing and
visualising data from any parameter estimation code. The
easy to use Python executable scripts and extensive
online documentation has resulted in PESummary becoming a key
component in the international gravitational-wave
analysis toolkit. PESummary has been developed to be more than just a post-processing
tool with all outputs fully self-contained. PESummary has become central to making gravitational-wave inference analysis open and easily reproducible.
\end{abstract}

\begin{keyword}
Parameter Estimation \sep Python \sep html \sep javascript \sep software



\end{keyword}

\end{frontmatter}

%% file: introduction.tex
\section{Motivation and significance}
\label{section:introduction}

Bayesian inference is central for many areas of science~\cite[e.g. ][]{jackman2009bayesian, ellison2004bayesian, beaumont2004bayesian, trotta2008bayes} as it allows for the identification of model parameters which best describes the collected data~\cite[e.g. ][]{gelman2013bayesian, harney2016bayesian}, see Section~\ref{sec:tables} for more details. Since the first detection of gravitational-waves (GWs)~\cite{Abbott:2016blz}, cosmic ripples predicted by Einstein's theory of General Relativity~\cite{Einstein:1916cc, Einstein:1918btx}, Bayesian inference has been used to infer the astrophysical source properties from GWs measurements~\cite[see e.g. ][]{TheLIGOScientific:2016wfe, Abbott:2018wiz, abbott2019gwtc}, understanding the properties of noise within the GW detectors~\cite{cornish2015bayeswave} and understanding the astrophysical population of GW sources~\cite{abbott2016binary}. For a detailed review of how Bayesian inference is used within the GW community see~\cite{Thrane:2018qnx}. In addition, with more areas of science entering the `open data era', there is a need for a robust, easy to use and code agnostic software to interpret and distribute the output from all Bayesian inference codes.
\par

\begin{figure}[t!]
    \centering
    \includegraphics[width=0.8\textwidth]{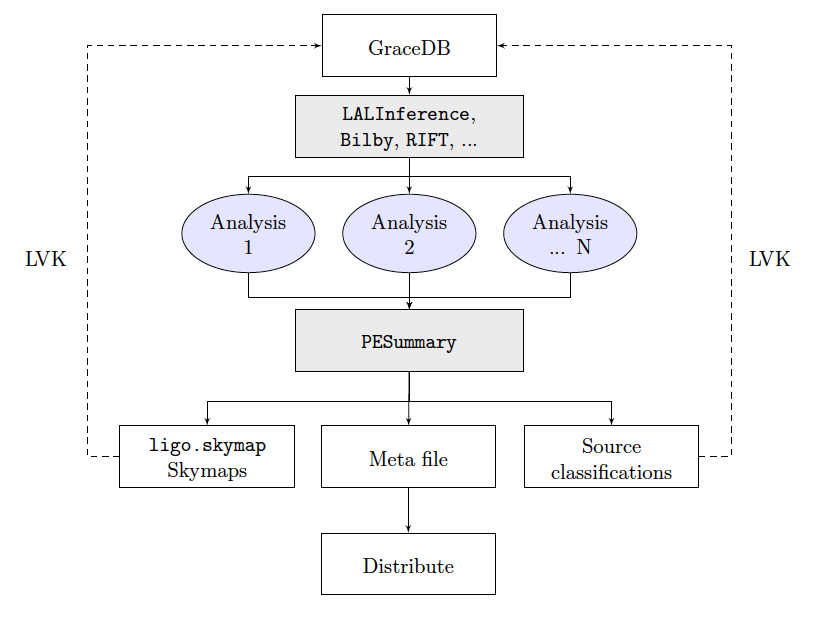}
    \caption{Flow chart showing the role of PESummary within the gravitational wave community. Dashed paths are specific for the LIGO Scientific, Virgo and KAGRA collaborations (LVK).
    Once a gravitational-wave is observed and uploaded to GraceDB, numerous Bayesian inference analyses are launched to extract the astrophysical source properties from the gravitational-wave measurement. The output data from all analyses is then passed to {\code{PESummary}} for post-processing. Alongside webpages for displaying, interpreting and comparing the Bayesian inference data, {\code{PESummary}} produces skymaps and source classifications, and a single universal `metafile' containing information about each Bayesian inference analyses  undertaken. This file can then distributed to the wider community. For the LVK workflow, skymaps and source classification probabilities are uploaded to GraceDB for circulation.}
    \label{fig:pesummary_with_gw_workflow}
\end{figure}

PESummary, the Parameter Estimation Summary page builder, is a
Python package that provides an intuitive, object-oriented user
interface for displaying, interpreting, comparing and distributing
posterior samples from any parameter estimation code. PESummary's
unified input/output (I/O) system enables the comparison of samples
from codes that previously stored data in incompatible formats.
\par
Since its first use in 2019, PESummary has grown to be a key component
in the workflow of the LIGO Scientific~\cite{LIGO}, Virgo~\cite{VIRGO} and KAGRA~\cite{KAGRA} collaborations (LVK), as shown in Figure~\ref{fig:pesummary_with_gw_workflow}. It is  used to analyse and compare the outputs from the popular GW Bayesian inference software packages:
{\code{LALInference}}~\cite{veitch2015parameter}, {\code{RIFT}}~\cite{Pankow:2015cra, Lange:2017wki, Wysocki:2019grj}, {\code{PyCBC}}~\cite{Biwer:2018osg} and
{\code{Bilby}}~\cite{ashton2019bilby, bilbygwtc, Smith:2019ucc} (both {\code{PyCBC}} and {\code{Bilby}} provide an interface for a number of popular sampling packages~\cite[e.g.][]{ForemanMackey:2012ig, speagle2020dynesty}), as well as circulating both skymaps through the
GW candidate event database GraceDB~\cite{gracedb} and Bayesian inference data
through the LIGO Document Control Center.
PESummary was critical in the parameter estimation analysis of
GW190412~\citep{ligo2020gw190412}, GW190425~\citep{Abbott:2020uma} and the
re-analysis of the first gravitational-wave transient catalog
(GWTC-1)~\citep{abbott2019gwtc} using
{\sc{Bilby}}. The released data
were also in the PESummary data format~\citep{data:GW190412, data:GW190425}.
\par
This article does not present a complete record of all the capabilities
of PESummary, for more information, see \href{https://lscsoft.docs.ligo.org/pesummary/stable_docs/index.html}{https://lscsoft.docs.ligo.org/\linebreak pesummary/stable\_docs/index.html}.

%% file: software.tex
\section{Software description}
\label{section:software}

The main PESummary interface is implemented in pure Python~\citep{python}
with the core library relying on a number of established scientific
programming packages~\citep{numpy, collette_python_hdf5_2014, Hunter:2007, corner, mckinney-proc-scipy-2010, astropy:2013, astropy:2018, da2019tqdm, plotly} and the GW library
requiring custom GW data analysis software: {\code{LALSuite}}~\citep{lalsuite}
and {\code{GWpy}}~\citep{gwpy}.
\par
PESummary can be used for more than simply post-processing parameter estimation results. This library includes, but not limited to, new Kernel Density Estimators~\cite{silverman1986density} to estimate PDFs of random variables within specified domains (one and two dimensional), a unified infrastructure for reading data in different formats, a release ready {\code{HDF5}}~\cite{collette_python_hdf5_2014} file for storing one or more Bayesian inference analyses, a comprehensive plotting suite and webpage module for analysing and comparing Bayesian inference data, and checkpointing, a requirement for cloud based computing.

PESummary is designed to be as modular and
adaptable as possible, ensuring the code will age gracefully with advances
in Bayesian inference software. Where possible, the code is kept general meaning additional post-processing techniques can easily be implemented into the PESummary framework.

\subsection{Software Architecture}

PESummary is structured around a small number of class objects. Each object
provides class and instance methods to provide the user with a complete
interface for all operations.

\subsubsection{Tabular Data}
\label{sec:tables}

Bayesian inference returns the best fit model parameters through a \emph{posterior} probability distribution function (PDF) calculated through Bayes' theorem~\cite{joyce2003bayes},

\begin{equation}
    p(\theta | d) = \frac{p(\theta)p(d | \theta)}{p(d)}
\end{equation}
where $p(\theta | d)$ is the posterior probability of the model having parameters $\theta$ given the data $d$, $p(\theta)$ is the prior probability of the model having parameters $\theta$, $p(d | \theta)$ is the likelihood of observing the data given a model with parameters $\theta$ and $p(d)$ is the evidence of the data, a model independent quantity. Often the likelihood and the posterior probability distribution function are unknown analytically. Consequently, most Bayesian inference packages are designed to draw samples from the unknown posterior PDF~\cite{metropolis1949monte, skilling2006nested}. Samples approximating a posterior PDF for a single parameter, otherwise known as a marginalized posterior PDF, are then identified by integrating the posterior PDF over all other parameters. Often, these posterior samples are output in the form of a table saved in a `result file'.

PESummary's stores posterior samples in a custom high level table structure, {\code{pesummary.utils.samples\_dict.SamplesDict}}, a subclass
of the Python builtin {\code{dict}} object. This class offers numerous methods for analysing the stored data, including the {\code{.plot}} method.  Each marginalized posterior distribution
is stored as a {\code{pesummary.utils.samples\_dict.Array}} object, an inherited class of {\code{numpy.ndarray}}~\citep{walt2011numpy}. This
structure provides direct access to the optimised array functions from
NumPy~\citep{numpy} and the usability and familiarity of dictionaries.
\par
A popular algorithm for sampling from the posterior distribution is Markov-Chain Monte-Carlo~\cite{metropolis1949monte}. For this case, multiple Markov chains are often run in parallel to test convergence through the Gelman-Rubin statistic~\citep{brooks1998general}. PESummary stores multiple Markov chains in the {\code{pesummary.utils.samples\_dict.MCMCSamplesDict}}) class, a dictionary where each chains posterior samples are represented by a {\code{pesummary.utils.samples\_dict.SamplesDict}} object. Algorithms for removing burnin are available via the {\code{.burnin()}}
method. Convergence between chains can be measured via the {\code{.gelman\_rubin()}} method.
\par
Often multiple Bayesian inference analyses are performed to identify how the PDFs vary for different settings. Consequently, PESummary provides the {\code{pesummary.utils.samples\_dict.MultiAnalysisSamplesDict}} \newline class, inherited from the the Python builtin {\code{dict}} object, for storing multiple {\code{pesummary.utils.samples\_dict.SamplesDict}} tables, each with an assigned label.

\subsection{Packages}

PESummary has followed the same methodology as {\sc{Bilby}}
by separating the top level code into 2 packages: {\code{core}} and {\code{gw}}.
The {\code{core}} package provides all of the necessary code for analysing,
displaying and comparing data files from general inference problems. The {\code{gw}}
specific package contains GW functionality, including converting posterior
distributions, deriving event classifications and GW specific plots, see
Sec.\ref{sec:gwpackage}.

\subsubsection{The {\code{core}} package}

The {\code{pesummary.core}} package provides all of the code required for 
post-processing general inference result files. It is designed to be generic and therefore work with the output from any Bayesian inference software. It provides a unified
interface for producing plots, calculating useful statistics and generating webpages.

Plots are generated via the {\code{pesummary.core.plots.plot}} module. One dimensional marginalized posterior distributions are visualised as either a histogram, a kernel density estimate ({\code{scipy.stats.gaussian\_kde}} or {\code{pesummary.core.plots.plots.bounded\_1d\_kde.Bounded\_1d\_kde}}) or cumulative distributions. Diagnostic plots displaying the marginalized trace or autocorrelation are also available. Comparison plots between multiple result files can be achieved by producing comparison one dimensional marginalized posterior distributions, cumulative distributions, box plots, violin plots, two dimensional contour and scatter plots or Jenson--Shannon~\citep{61115} and Kolmogorov--Smirnov~\citep{kolmogorov1933, smirnov1948} statistics.

The {\code{pesummary.core.webpage.webpage}} module is a Python wrapper for writing HTML. The {\code{pesummary.core.webpage.webpage.page}} class provides functionality for generating multi-level navigation bars, tables of images and/or data, modal carousels and more. The design and functionality of the webpages are controlled through the {\code{pesummary.core.css}} and {\code{pesummary.core.js}} modules; each containing custom style sheets (CSS) or JavaScript files (JS) respectively.
\par
The {\code{combine\_corner.js}} script contains functionality to generate custom corner plots by manipulating a pre-made figure~\cite[made with e.g. ][]{corner}. By providing an interface for the user to specify parameters they wish to compare, the PESummary webpages allow for the PDF, and its correlations, to be interactively explored.

\subsubsection{The {\code{gw}} package}\label{sec:gwpackage}

The {\code{gw}} package provides the functionality for parameter estimation
specific to GW astronomy. Building on the {\code{core}} package,
the {\code{gw}} module provides additional methods and classes for handling GW
specific data files. Although the {\code{gw}} package is tailored for compact binary coalescence data files, we provide GW specific methods which can be applied to the Bayesian inference data from any transient GW source.
\par
The {\code{gw}} package provides a comprehensive conversion module \linebreak
({\code{pesummary.gw.file.conversions}}) for deriving alternative
posterior distributions from the input, e.g. the primary mass and secondary mass from chirp mass and mass ratio. Assuming a binary black hole merger,
the conversion class provides multiple methods for estimating the properties
of the final black hole: final mass (or equivalently the energy radiated in
gravitational-waves), final spin, peak luminosity and final kick
velocity~\citep{Babak:2016tgq, Ossokine:2020, Husa:2015iqa, Khan:2018fmp, Khan:2019kotf, varma2019surrogate, varma2019surrogateprecessing, Hofmann:2016yih, spinfit-T1600168, Healy:2016lce, Jimenez-Forteza:2016oae, Keitel:2016krm, Varma:2020kick}. All fits are calibrated to numerical relativity
and are interchangeable via keyword arguments provided to the conversion class.
\par
On the 17th August 2017, alerts~\cite{GW170817_GCNs} were sent out to more than 60 international groups of astronomers notifying them that a GW had just been detected: GW170817~\cite{TheLIGOScientific:2017qsa}. Each group began observing the night sky to independently observe the source of the GW. GW170817 opened the window of a long-awaited multi-messenger astronomy~\cite{GBM:2017lvd}. Through interfacing with the {\code{ligo.skymap}}~\citep{ligoskymap} package, PESummary provides an intuitive method for generating skymaps: data files showing the most likely source location of the GWs. These skymaps are distributed to astronomers through GraceDB, see Figure~\ref{fig:pesummary_with_gw_workflow}. In the last GW observing period, 26 alerts were sent out regarding possible GW candidates, each containing skymaps~\cite{O3a_GCNs}. PESummary also interfaces with the {\code{PEPredicates}}~\citep{PEPredicates} and {\code{P-Astro}}~\citep{PAstro} packages to provide source classification probabilities for the type of compact binary merger observed; such as the probability that the system is a binary black hole or a binary neutron star. Both the skymap and source classification probabilities are vital for electromagnetic follow-up campaigns.
\par
PESummary takes advantage of the {\code{plotly}}~\citep{plotly} interactive plotting
package to produce interactive corner plots for extrinsic and intrinsic parameters. The
{\code{pesummary.gw.plots.publication}} module also allows for `publication'
quality plots to be produced, for instance those in~\citep{abbott2019gwtc}.

\subsection{Software Functionalities}

\subsubsection{Unified input/output}\label{sec:read}

\begin{table}[t!]
\begin{tabular}{|l|l|p{4.0cm}|p{4.0cm}|}
\hline
Package & Format & Class & Description \\
\hline
{\code{core}} & .dat & \code{.default.Default} & `.dat' format with parameter names as header \\
\hline
{\code{core}} & .txt & \code{.default.Default} & `.txt' format with parameter names as header \\
\hline
{\code{core}} & .h5/.hdf5 & \code{.default.Default} & `.hdf5' format with data stored in a group called `posterior' or `posterior\_samples'  \\
\hline
{\code{core}} & .json & \code{.default.Default} & `.json' format with posterior samples stored in a group called `posterior' or `posterior\_samples'  \\
\hline
{\code{core}} & {\sc{Bilby}} & \code{.bilby.Bilby} & Result file produced by {\sc{Bilby}}  \\
\hline
{\code{gw}} & {\sc{Bilby}} & \code{.bilby.Bilby} & Result file produced by {\sc{Bilby}} \\
\hline
{\code{gw}} & {\sc{LALInference}} & \code{.lalinference.\linebreak LALInference} & Result file produced by {\sc{LALInference}} \\
\hline
\end{tabular}
\caption{A selection of formats that can be read in with PESummary and accessible
through the unified I/O interface for the listed class object(s). The core classes
are all prepended by {\code{pesummary.core.file.formats}}, the gw classes are all
prepended by {\code{pesummary.gw.file.formats}}.}
\label{tab:formats} 
\end{table}

Bayesian inference packages output their posterior samples in varying formats. In just the GW community alone, {\code{LALInference}}, {\code{Bilby}} and {\code{RIFT}} output their data in 3 different formats: {\code{HDF5}}~\cite{collette_python_hdf5_2014}, {\code{JSON}}~\cite{JSON} and {\code{dat}}. Although {\code{Bilby}} is able to output it's data in {\code{HDF5}}, it's posterior samples are stored differently to {\code{LALInference}}: {\code{LALInference}} stores posterior samples as a {\code{numpy.recarray}} while {\code{Bilby}} writes each marginalized posterior distribution as a separate dataset, and an alternative software package for reading {\code{HDF5}} files is required. Each data file also stores different metadata in different locations. This incompatibility makes it difficult to compare the contents from different Bayesian inference samplers both in and out of the GW community.

PESummary provides a unified infrastructure for reading data in different formats via the {\code{pesummary.io}} module. The universal {\code{pesummary.io.read()}} function reads data stored in all common file formats
via the {\code{pesummary.core.file.formats}} module. The GW specific package
extends the allowed file formats via the {\code{pesummary.gw.file.formats}} module. See
Table~\ref{tab:formats} for a reduced list. Once read, the posterior samples from different data files can be compared through the common {\code{.samples\_dict}} property. All read result files can be written to a specified file format via the common {\code{.write}} method. This method calls the universal {\code{pesummary.io.write()}} function.
\par
PESummary offers the {\code{pesummary.core.file.meta\_file}} module for producing a  universal file containing posterior samples and metadata for one or more analyses. This single `metafile' aims to store all information regarding the Baysian inference analysis. This includes prior samples, configuration files, injection information, environment information for reproducibility etc. For GW analyses, the {\code{pesummary.gw.file.meta\_file}} module allows for the PSD, skymap data and calibration information to also be saved. 

\subsubsection{Executables}

\begin{table}[t!]
\begin{tabular}{|l|p{8.5cm}|}
\hline
Executable & Description \\
\hline
{\code{summaryclassification}} & Generate GW event classification probabilities\\
\hline
{\code{summaryclean}} & Clean and convert an input result file \\
\hline
{\code{summarycombine}} & Combine multiple result files into a single format \\
\hline
{\code{summarycompare}} & Compare the contents of multiple files \\
\hline
{\code{summarydetchar}} & Generate GW Detector characterisation plots \\
\hline
{\code{summarygracedb}} & Retrieve data from the online GW Candidate Event Database~\citep{gracedb} \\
\hline
{\code{summarymodify}} & Modify a PESummary meta file \\
\hline
{\code{summarypages}} & Generate webpages, plots and a meta file for N result files \\
\hline
{\code{summarypageslw}} & Generate webpages, plots and meta files for a select number of parameters \\
\hline
{\code{summaryplots}} & Generate specific plots for a given result file \\
\hline
{\code{summarypipe}} & Generate a {\code{summarypages}} executable given a GW run directory \\
\hline
{\code{summarypublication}} & Generate publication plots for N result files \\
\hline
{\code{summaryrecreate}} & Launch the GW analysis that was used to generate the PESummary meta file \\
\hline
{\code{summaryversion}} & Version of PESummary installed \\
\hline
\end{tabular}
\caption{A selection of executable python scripts provided by PESummary}
\label{tab:executables} 
\end{table}

\begin{figure}
    \centering
    \includegraphics[width=\textwidth]{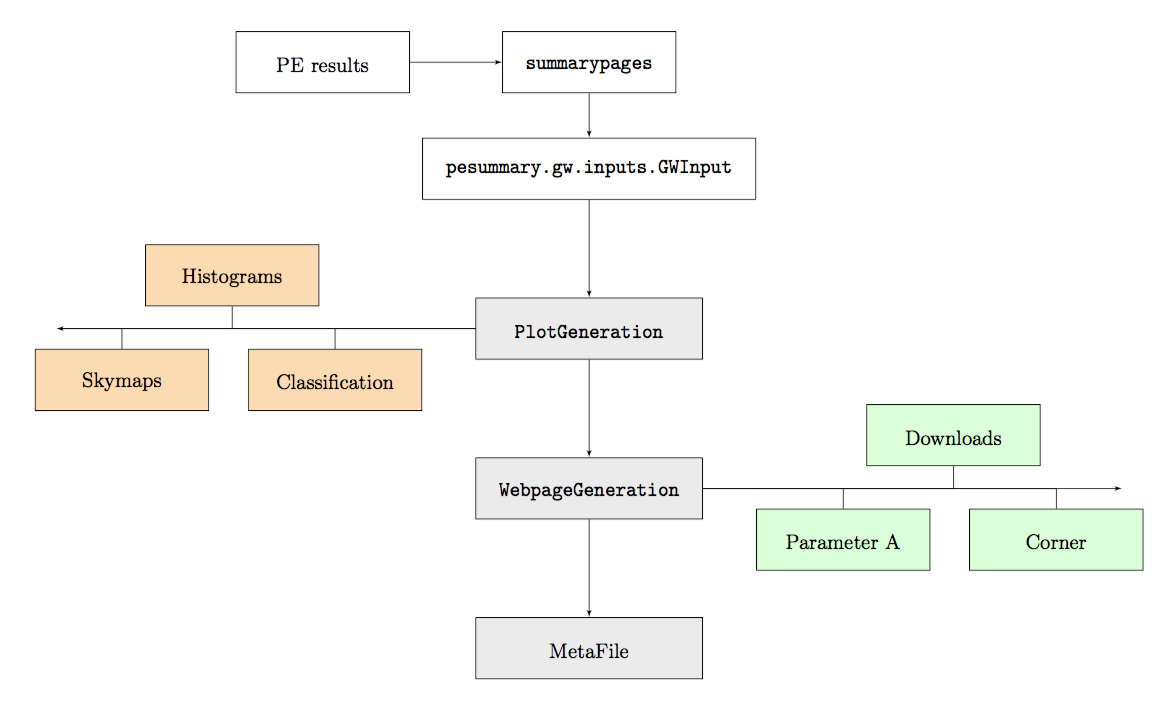}
    \caption{Workflow for the most general executable provided by PESummary: {\code{summarypages}}. Here we show the workflow when the {\code{gw}} package is used.}
    \label{fig:summarypages_workflow}
\end{figure}

PESummary provides multiple executable Python scripts to act as an intermediary
between the command line and the core functionality. See Table~\ref{tab:executables}.
{\code{summarypages}} is the main executable and combines {\code{summaryclassification}},
{\code{summaryclean}}, {\code{summarycombine}}, {\code{summarydetchar}}, {\code{summaryplots}}
and {\code{summarypublication}}. {\code{summarypages}} is the most general executable
provided by PESummary. Its GW workflow is described in Fig.~\ref{fig:summarypages_workflow}. The core workflow is similar, except GW specific plots and webpages are not generated and the core classes are used.
\par
The {\code{summarypages}} executable takes one or more result files as input via the {\code{--samples}} command line argument. A plethora of optional command line arguments are also available to allow for complete customisation\footnote{For details run {\code{summarypages --help}} or vist the online documentation available \href{https://lscsoft.docs.ligo.org/pesummary/stable_docs/core/cli/summarypages.html}{https://lscsoft.docs.ligo.org/pesummary/stable\_docs/core/cli/summarypages.html}.}. All options are then passed to the {\code{pesummary.gw.inputs.GWInput}} class (inherited from the {\code{pesummary.core.inputs.Input}} class) which checks the inputs. Amongst these checks, the samples are extracted from all result files (see Section~\ref{sec:read}), converted to generate additional PDFs, and stored in a {\code{pesummary.utils.samples\_dict.MultiAnalysisSamplesDict}} object with an assigned label.
\par 
All properties of the {\code{GWInput}} class are then used to initiate the {\code{pesummary.cli.summaryplots.PlotGeneration}} class. Here, all plots are generated on multiple CPUs, if specified, with the {\code{.generate\_plots}} method. This includes source classification plots, marginalized posteriors, interactive plots and more. Initial skymaps are generated based on a two-dimensional histogram of the right ascension and declination samples. A subprocess is then launched to generate the more accurate {\code{ligo.skymap}} skymap. This implementation allows for the rest of the pipeline to continue without having to wait for the often time-consuming {\code{ligo.skymap}} skymap to be produced. Once complete, the initial skymap is overwritten.
\par
All webpages are then produced by running the {\code{.generate\_webpages}} method once the {\code{pesummary.cli.summarypages.WebpageGeneration}} class is initialised with all {\code{GWInput}} properties. This includes single webpages for each parameter in each result file, interactive corner pages, comparison pages for all common parameters, a version page providing version and environment information and a downloads page. Finally a single metafile containing all information from the run is produced. This includes environment information, posterior samples, command line arguments and more. This file is available for download via the downloads page.
\par
The output from {\code{summarypages}} is completely self-contained; allowing for it to distributed if required. An example of the {\code{summarypages}} output can be seen here: \href{https://pesummary.github.io/GW190412/home.html}{https://pesummary.github.io/GW190412/home.html}. This output was produced from Listing~\ref{summarypages:GW190412}.
\par
Amidst the collection of optional command line arguments, the {\code{pesummary.gw}} module also provides a dynamic argument parser. This allows for dependent arguments to be passed from the command line (see section~\ref{sec:dynamic}).

%% file: examples.tex
\section{Illustrative Examples}
\label{}

Below we provide a limited set of examples to demonstrate some of the features of PESummary. All data and scripts that are used as part of this section can be downloaded from \href{https://github.com/pesummary/pesummary-paper-data}{https://github.com/pesummary/pesummary-paper-data}. More examples can be found in the PESummary repository: \href{https://git.ligo.org/lscsoft/pesummary/-/tree/master/examples}{https://git.ligo.org/lscsoft/pesummary/-/tree/master/examples}. The following examples assume that you have cloned the \href{https://github.com/pesummary/pesummary-paper-data.git}{pesummary-paper-data repository} and you are in the base directory.

\subsection{Example 1: Running with {\code{emcee}}}

\begin{figure}[t!]
    \centering
    \includegraphics[width=0.49\textwidth]{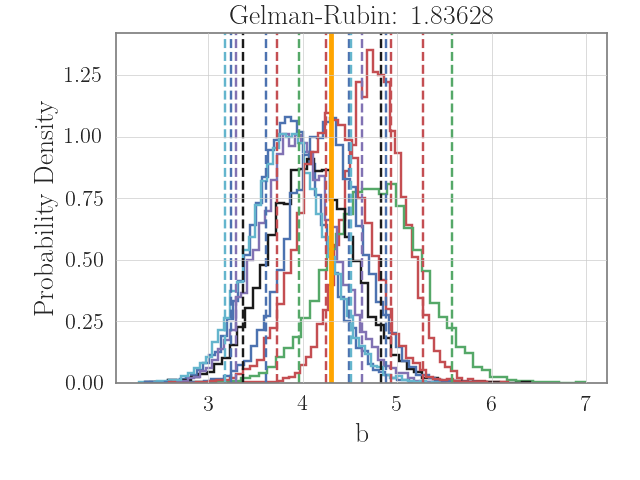}
    \includegraphics[width=0.49\textwidth]{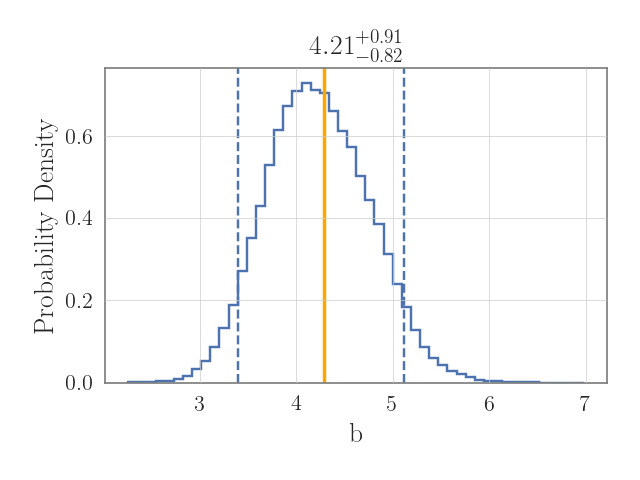}
    \caption{The output marginalized posterior for each chain (Left) and the posterior from combining all chains (Right) resulting from Listing~\ref{list:emcee}. The injected value is shown in orange.}
    \label{fig:emcee}
\end{figure}

This example shows the flexibility of the PESummary post-processing package. We
run the \href{https://emcee.readthedocs.io/en/stable/tutorials/line/}{\textit{Fitting a
model to data}} tutorial provided as part of the {\code{emcee}}\citep{2013PASP..125..306F} sampling
package. We save the posterior samples to a `.dat' file and run with 8 chains.
All post-processing is then handled with PESummary, see Figure~\ref{fig:emcee} for example output. The {\code{emcee}} code snippet, as well as the posterior samples can be downloaded from the \href{https://github.com/pesummary/pesummary-paper-data}{pesummary-paper-data repository}.

\begin{minipage}{.9\textwidth}
  \begin{lstlisting}[%
    caption={Running PESummary on the {\code{emcee}} output.},
    label={list:emcee},
    language=bash,
  ]

python emcee_tutorial.py
chains=($(ls emcee_output/chain_*.dat))
summarypages --webdir ./webpage --samples ${chains[@]}
             --labels tutorial --mcmc_samples --inj_file emcee_output/injected.txt
  \end{lstlisting}
\end{minipage}

\subsection{Example 2: Reading a result file}

This example demonstrates how to read in a PESummary `metafile' using the {\code{pesummary.io}} module. We use the {\code{posterior\_samples.h5}} file produced as part of Listing~\ref{list:emcee} (running Listing~\ref{list:emcee} is not necessary for this example, the metafile is available in the \href{https://github.com/pesummary/pesummary-paper-data.git}{pesummary-paper-data repository}) and specify that we would like to use the {\code{core}} package meaning only the core file formats are allowed. We show how to extract the posterior samples for a specified analysis, how to access the marginalized posterior samples, and how to extract the prior samples and any stored configuration files. Some of the attributes of each object are also shown. For full documentation, run the inbuilt Python help function {\code{help}}~\cite{python-help}.

\begin{minipage}{.9\textwidth}
  \begin{lstlisting}[%
    caption={Example code to read in a PESummary result file with PESummary.},
    label={code:asd},
    language=Python,
  ]
from pesummary.io import read

f = read("posterior_samples.h5", package="core")
multi_analysis_samples = f.samples_dict
print(type(multi_analysis_samples))
analysis = multi_analysis_samples.labels[0]
posterior_samples = multi_analysis_samples[analysis]
print(type(posterior_samples))
print(posterior_samples.keys())
marginalized = posterior_samples["m"]
print(type(marginalized))
print(marginalized.average(type="median"))
priors = f.priors["samples"]
config = f.config
  \end{lstlisting}
\end{minipage}

\subsection{(GW) Example 3: analysing public LIGO and Virgo posterior samples}

\begin{figure}[h!]
    \centering
    \includegraphics[width=0.6\textwidth]{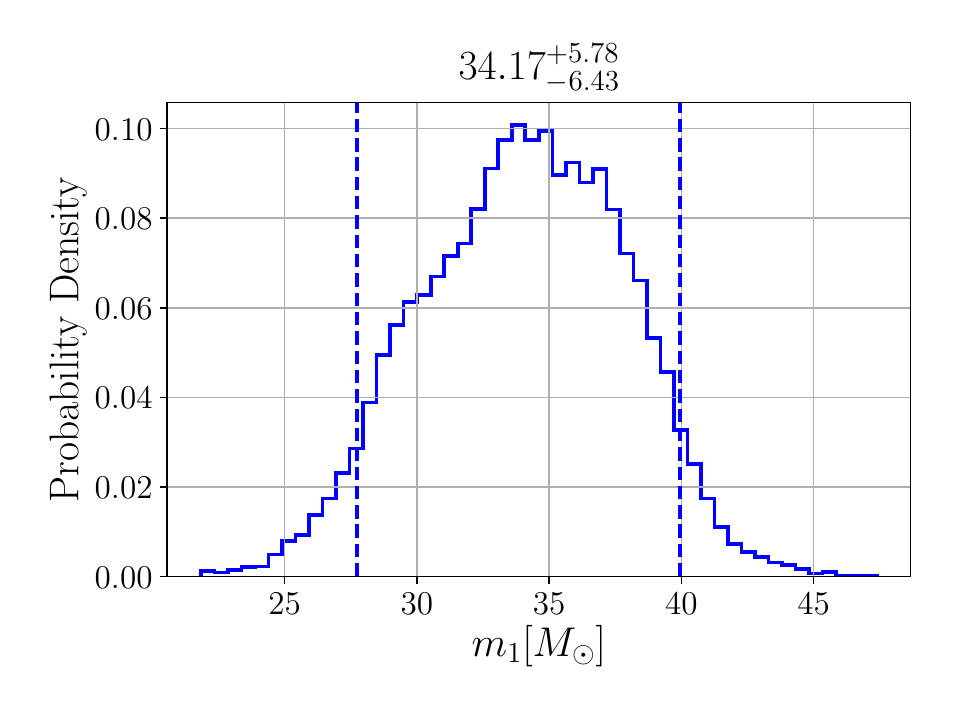}
    \includegraphics[width=0.6\textwidth]{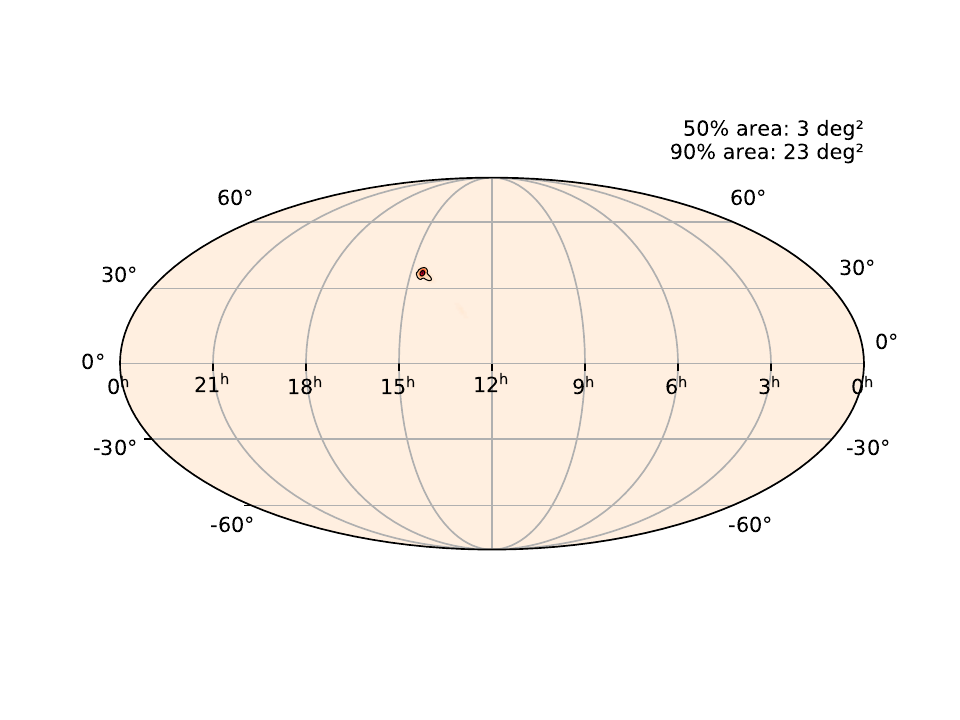}
    \includegraphics[width=0.6\textwidth]{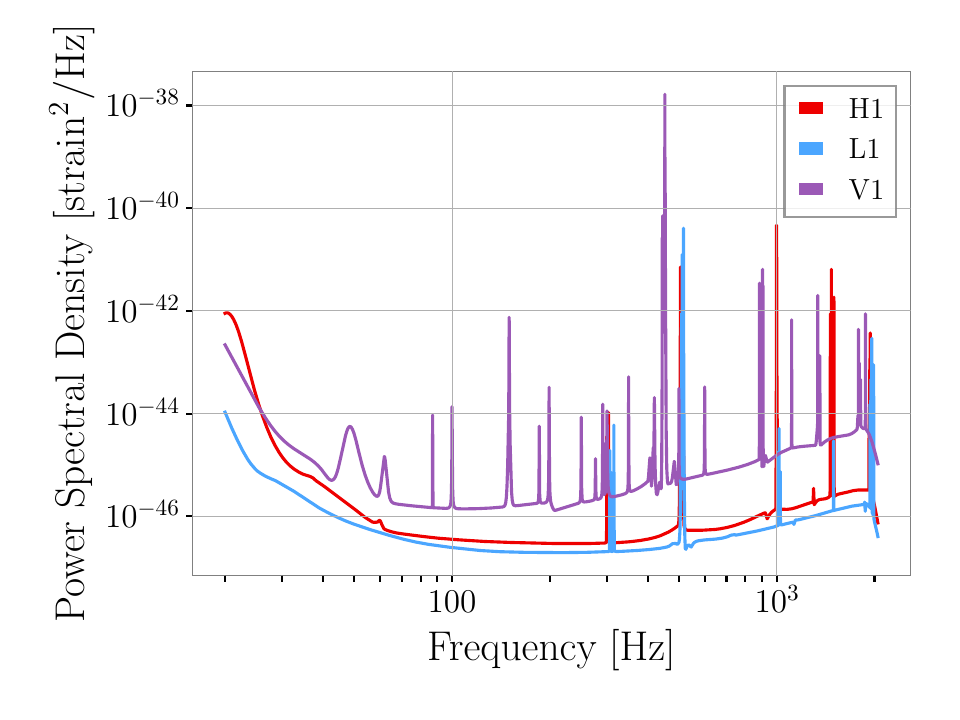}
    \caption{The output marginalized posterior (top), skymap (middle) and PSD plot (bottom) from Listing~\ref{list:GW190412_read}.}
    \label{fig:GW190412}
\end{figure}

This example demonstrates how to extract and analyse public LIGO and Virgo posterior samples. It includes demonstrations of how to produce `standard' plots for the `combined' analysis stored in the PESummary `metafile' through the {\code{.plot}} method, see Figure~\ref{fig:GW190412}. We use the publicly available \href{https://dcc.ligo.org/DocDB/0163/P190412/009/posterior\_samples.h5}{https://dcc.ligo.org/DocDB/0163/P190412/009/posterior\_samples.h5} file, which has been copied to the \href{https://github.com/pesummary/pesummary-paper-data}{pesummary-paper-data repository}. This specific file was chosen since it includes additional information such as the PSD, calibration envelopes, priors etc. Owing to GitHub’s strict 100MB file limit, the file located in the pesummary-paper-data repository is stored under {\code{git-lfs}}\cite{gitlfs}. For instructions on how to successfully download this file, see the \href{https://github.com/pesummary/pesummary-paper-data/blob/main/README.md}{pesummary-paper-data README}.

\begin{minipage}{.9\textwidth}
  \begin{lstlisting}[%
    caption={Extracting data from public LIGO and Virgo posterior samples},
    label={list:GW190412_read},
    language=Python,
  ]
from pesummary.io import read

f = read("GW190412_posterior_samples.h5", package="gw")
analysis = "combined"
posterior_samples = f.samples_dict[analysis]
psds = f.psd[analysis]
calibration_envelope = f.priors["calibration"][analysis]
prior_samples = f.priors["samples"][analysis]

hist = posterior_samples.plot("mass_1", type="hist")
hist.show()
skymap = posterior_samples.plot(type="skymap")
skymap.show()
plot = psds.plot(fmin=20)
plot.show()
  \end{lstlisting}
\end{minipage}

\subsection{(GW) Example 4: Producing a summary page for public LIGO and Virgo posterior samples}

This example demonstrates how a summary page can be generated
from publicly available LIGO and Virgo data files. We use the same publicly available data file as Listing~\ref{list:GW190412_read}. This `metafile' contains a total of 10 Bayesian inference analyses, each using different standard models used by the LVK. In this example, we choose to compare only a select subset: an analysis conducted with the IMRPhenomPv3HM model~\cite{Khan:2019kot} and the SEOBNRv4PHM model~\cite{Ossokine:2020kjp, Pan:2013rra, Babak:2016tgq} (see~\cite{LIGOScientific:2020stg} for details). We specify that we would like to use the {\code{gw}} package ({\code{--gw}}), run on 15 CPUs ({\code{--multi\_process 15}}), and generate `publication' quality plots ({\code{--publication}}). The output page can be found here: \href{https://pesummary.github.io/GW190412/home.html}{https://pesummary.github.io/GW190412/home.html}.

\begin{minipage}{.9\textwidth}
  \begin{lstlisting}[%
    caption={Generating a html page to compare and analyse the public LIGO and Virgo posterior samples. This may take some time.},
    label={summarypages:GW190412},
    language=bash,
  ]
summarypages --webdir ./GW190412 \ 
             --samples GW190412_posterior_samples.h5 \ 
             --gw --compare_results IMRPhenomPv3HM SEOBNRv4PHM \ 
             --publication \ 
             --multi_process 15
  \end{lstlisting}
\end{minipage}

\subsection{(GW) Example 5: PESummary's dynamic argument parser}
\label{sec:dynamic}

{\code{argparse}}, the Python module for handling command-line arguments (CLA)~\cite{argparse}, requires a known list of arguments. This means that they cannot depend on another variable. Through the \code{pesummary.gw.command\_line} module, PESummary allows the user to specify CLAs which change depending on the provided label. Below we show how PSDs can be provided to {\code{summarypages}} dynamically through the {\code{--\{\}\_psd}} CLA. Result files and psd data was created using the `make\_data\_for\_listing5.py' script made available in the \href{https://github.com/pesummary/pesummary-paper-data}{pesummary-paper-data repository}.

\begin{minipage}{.9\textwidth}
  \begin{lstlisting}[%
    caption={Example usage of PESummary's dynamic argument parser},
    label={list:argparse},
    language=bash,
  ]
summarypages --webdir ./webpage \ 
             --samples test.hdf5 test.json \ 
             --labels hdf5_example json_example \ 
             --hdf5_example_psd H1:psd_H1.dat \ 
             --json_example_psd V1:psd_V1.dat L1:psd_L1.dat \ 
             --gw
  \end{lstlisting}
\end{minipage}

\subsection{(GW) Example 6: Reproducing LIGO and Virgo plots}

\begin{figure}[t!]
    \includegraphics[width=0.5\textwidth]{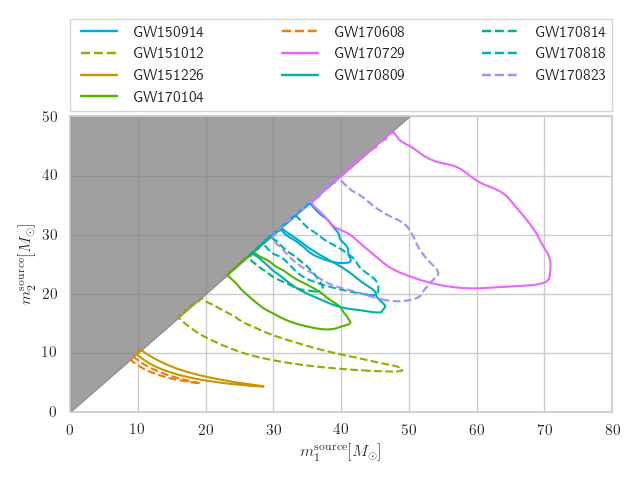}
    \includegraphics[width=0.5\textwidth]{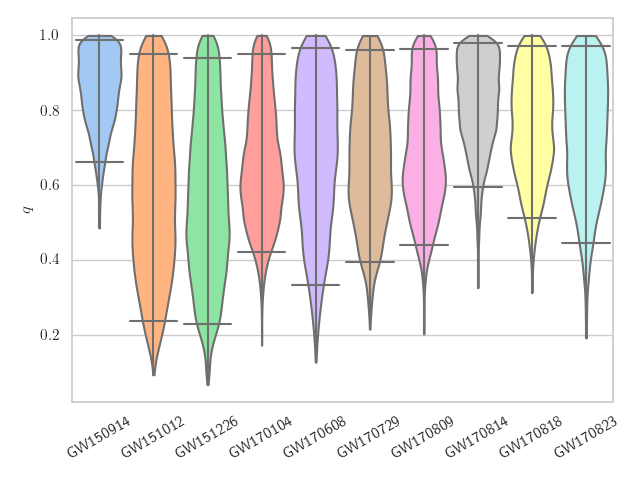}
    \caption{Two plots output from Listing~\ref{summarypublication:GW190412}. Left: Recreation of Figure 4 in Ref.~\citep{abbott2019gwtc}, Right: Recreation of Figure 5 in Ref.~\citep{abbott2019gwtc}.}
    \label{fig:gwtc1}
\end{figure}

This example demonstrates how to reproduce a subset of plots in the first GW transient catalog (GWTC-1)~\citep[Figures~4 and 5 in Ref.][]{abbott2019gwtc}. We use publicly available posterior samples released as part of GWTC-1 (ignoring the GW170809~\cite{abbott2019gwtc} prior choices file): \href{https://dcc.ligo.org/public/0157/P1800370/005/GWTC-1\_sample\_release.tar.gz}{https://dcc.ligo.org/public/0157/P1800370/005/GWTC-1\_sample\_release.tar.gz}, which have been copied to the \href{https://github.com/pesummary/pesummary-paper-data}{pesummary-paper-data repository}. Figure~\ref{fig:gwtc1} shows an example of the output. When running this Listing, PESummary will print multiple warnings to {\code{stdout}}. These warnings are expected and shows that PESummary is robust to potential failures and will continue to produce an output while still warning the reader appropriately. One such example is a message warning the user that the `./GWTC-1\_sample\_release/GW170817\_GWTC-1.hdf5' cannot be read in and data will not be added to the plot as this file contains `multiple posterior sample tables: IMRPhenomPv2NRT\_highSpin\_posterior, IMRPhenomPv2NRT\_lowSpin\_posterior' and we have not specified which we wish to load from the command line.

\begin{minipage}{.9\textwidth}
  \begin{lstlisting}[%
    caption={Example code to generate GW plots with PESummary. This may take some time.},
    label={summarypublication:GW190412},
    language=bash,
  ]
FILES=($(ls ./GWTC-1_sample_release/*_GWTC-1.hdf5))

LABELS=()
for i in ${FILES[@]}; do
    label=`python -c "print('${i}'.split('_GWTC-1.hdf5')[0])"`
    LABELS+=($(python -c "print('${label}'.split('./GWTC-1_sample_release/')[1])"))
done

COLORS=('#00a7f0' '#9da500' '#c59700' '#55b300' '#f77b00' '#ea65ff' '#00b1a4' '#00b47d' '#00aec2' '#9f8eff')
LINESTYLES=(solid dashed solid solid dashed solid solid dashed dashed dashed solid)

summarypublication --plot 2d_contour \
                   --webdir ./GWTC-1_sample_release \
                   --samples ${FILES[@]} \
                   --labels ${LABELS[@]} \
                   --parameters mass_1_source mass_2_source \
                   --colors ${COLORS[@]} \
                   --linestyles ${LINESTYLES[@]} \
                   --publication_kwargs xlow:0 xhigh:80 ylow:0 yhigh:50

summarypublication --plot violin \
                   --webdir ./GWTC-1_sample_release \
                   --samples ${FILES[@]} \
                   --labels ${LABELS[@]} \
                   --parameters mass_ratio

  \end{lstlisting}
\end{minipage}

%% file: impact.tex
\section{Impact}
\label{section:impact}

PESummary is now a widely used library in the LVK. In
just over one year, PESummary has become the post-processing software for the main LVK
GW parameter estimation codes~\citep{ashton2019bilby, bilbygwtc, veitch2015parameter, smith2019expediting, Carullo:2019flw, Isi:2019aib, lange2018rapid}
and is relied upon for the open data release of the LVK parameter estimation
data products~\citep{publicdata}. Rather than releasing multiple data products in different
formats spread across numerous URLs, PESummary has greatly simplified this to simply releasing
a single file~\citep{data:GW190412, data:GW190425}. This can propel future research in the
field, as researches no longer have to create custom scripts for combining, retrieving, and
recreating the initial analysis.
\par
The flexibility and intuitive Python executables provided by PESummary, has significantly
improved the efficiency of researchers (often early-career graduates or graduate scientists),
by reducing the need for repetitive tasks. In particular, PESummary's interactive corner plots
has led to the increased knowledge of degeneracies between parameters for researchers within
the LVK.
\par
PESummary is also fully incorporated into the automated low-latency alert workflow~\citep{GWCelery}.
Posterior based classifications are therefore automatically produced and posted to the online GW
Candidate Event Database~\citep{gracedb} from the automatic parameter estimation follow-up. This
information will greatly aid electromagnetic followup campaigns.

%% file: conclusion.tex
\section{Conclusions}
\label{section:conclusion}

In this paper we have described PESummary, the Parameter Estimation Summary page builder. This Python package provides a modern interface for displaying
interpreting, comparing and distributing Bayesian inference data from any parameter estimation code. PESummary has been used extensively by the
international GW community, and has been crucial for Advanced LIGO's~\cite{TheLIGOScientific:2014jea} and Advanced Virgo's~\cite{acernese2014advanced} third gravitational wave observing run.

Although PESummary is primarily used for post-processing GW Bayesian inference data from compact binary coalescenses, looking forward, we plan to incorporate other GW fields, e.g. Tests of General Relativity~\cite[see e.g. ][]{abbott2019tests}. This will enable PESummary to be the driving force behind the post-processing and distribution of all Bayesian inference data output from the LVK. We will also continue to expand our already comprehensive plotting suite to improve its versatility for any Bayesian inference analysis.

%% file: conflict.tex
\section{Conflict of Interest}
We wish to confirm that there are no known conflicts of interest associated with this publication and there
has been no significant financial support for this work that could have influenced its outcome.

%% file: acknowledgements.tex
\section*{Acknowledgements}
\label{section:acknowledgements}

The authors wish to thank the reviewers for comments and suggestions concerning this manuscript. The authors would also like to thank the LIGO Scientific, Virgo and KAGRA collaboration for all constructive feedback, bug reports
and feature requests. A special thanks to Virginia D'Emilio, David Keitel,
Nicola De Lillo,  Nathan Johnson-McDaniel, and Philip Relton for reviewing PESummary. We wish to thank
Gregory Ashton, Christopher Berry, Sylvia Biscoveanu, Gregorio Carullo, Stephen Fairhurst, Edward Fauchon-Jones, Rhys Green, Carl-Johan Haster, Duncan Macleod, Sergeui Ossokine, Isobel Romero-Shaw,
Colm Talbot and John Veitch for useful discussions. We thank all past and current contributors; for a
full overview, see the \href{https://git.ligo.org/lscsoft/pesummary/-/graphs/master}{list of contributors
on Gitlab}. The authors are grateful for computational resources provided by the Leonard E Parker Center
for Gravitation, Cosmology and Astrophysics at the University of Wisconsin-Milwaukee and Cardiff University,
supported by National Science Foundation Grants PHY-1626190 and PHY-1700765 and STFC grant ST/I006285/1.